\documentclass[reprint,amsmath,amssymb,aip,rsi]{revtex4-1}

\usepackage{etoolbox}
\usepackage[utf8]{inputenc}
\usepackage[T1]{fontenc}  
\usepackage{graphicx}
\usepackage{dcolumn}
\usepackage{bm}
\usepackage{mathrsfs}
\usepackage{amsfonts}
\usepackage{textcomp}

\usepackage[table]{xcolor}  

\usepackage{mathptmx} 

\usepackage{upgreek}

\let\baraccent=\= 
\renewcommand{\=}[1]{\stackrel{#1}{=}} 

\usepackage{upgreek}

\DeclareRobustCommand\upmu{\ensuremath\mathrm\mu}

\makeatletter
\def\@email#1#2{%
 \endgroup
 \patchcmd{\titleblock@produce}
  {\frontmatter@RRAPformat}
  {\frontmatter@RRAPformat{\produce@RRAP{*#1\href{mailto:#2}{#2}}}\frontmatter@RRAPformat}
  {}{}
}%
\makeatother

\begin{document}

\title{\textbf{Cryogenic sensor enabling broad-band and traceable power measurements}}

\author{J.-P. Girard}
\affiliation{QCD Labs, QTF Centre of Excellence, Department of Applied Physics, Aalto University, P.O. Box 13500, FIN-00076 Aalto, Finland.}



\author{R.~E.~Lake}
\affiliation{Bluefors Oy, Arinatie 10, 00510 Helsinki, Finland}
\email{russell.lake@bluefors.com}

\author{W. Liu}
\affiliation{QCD Labs, QTF Centre of Excellence, Department of Applied Physics, Aalto University, P.O. Box 13500, FIN-00076 Aalto, Finland.}
\affiliation{IQM, Keilaranta 19, FI-02150 Espoo, Finland}

\author{R. Kokkoniemi}
\affiliation{QCD Labs, QTF Centre of Excellence, Department of Applied Physics, Aalto University, P.O. Box 13500, FIN-00076 Aalto, Finland.}
\affiliation{IQM, Keilaranta 19, FI-02150 Espoo, Finland}
\author{E. Visakorpi}
\affiliation{QCD Labs, QTF Centre of Excellence, Department of Applied Physics, Aalto University, P.O. Box 13500, FIN-00076 Aalto, Finland.}
\author{J. Govenius}
\affiliation{QCD Labs, QTF Centre of Excellence, Department of Applied Physics, Aalto University, P.O. Box 13500, FIN-00076 Aalto, Finland.}
\affiliation{VTT Technical Research Centre of Finland Ltd. \& QTF Centre of Excellence, P.O. Box 1000, 02044 VTT, Finland.}
\author{M. Möttönen}
\affiliation{QCD Labs, QTF Centre of Excellence, Department of Applied Physics, Aalto University, P.O. Box 13500, FIN-00076 Aalto, Finland.}
\affiliation{VTT Technical Research Centre of Finland Ltd. \& QTF Centre of Excellence, P.O. Box 1000, 02044 VTT, Finland.}


\date\today
\begin{abstract}
\noindent 
Recently, great progress has been made in the field of ultrasensitive microwave detectors, reaching even the threshold for utilization in circuit quantum electrodynamics. However, cryogenic sensors lack the compatibility with broad-band metrologically traceable power absorption measurements at ultralow powers, which limits their scope of applications. Here, we demonstrate such measurements using an ultralow-noise nanobolometer which we extend by an additional direct-current (dc) heater input. The tracing of the absorbed power relies on comparing the response of the bolometer between radio frequency (rf) and dc-heating powers traced to the Josephson voltage and quantum Hall resistance. To illustrate this technique, we demonstrate methods to calibrate the power that is delivered to the base temperature stage of a dilution refrigerator using our in-situ power sensor. As an example, we demonstrate the ability to measure accurately the attenuation of a coaxial input line between the frequencies of 50 MHz and 7 GHz with an uncertainty down to 0.1~dB at typical input power of -114 dBm.


\end{abstract}

\maketitle


\section{Introduction} 
\label{sec:intro}
\begin{figure*}[t]
    \includegraphics[width=\textwidth]{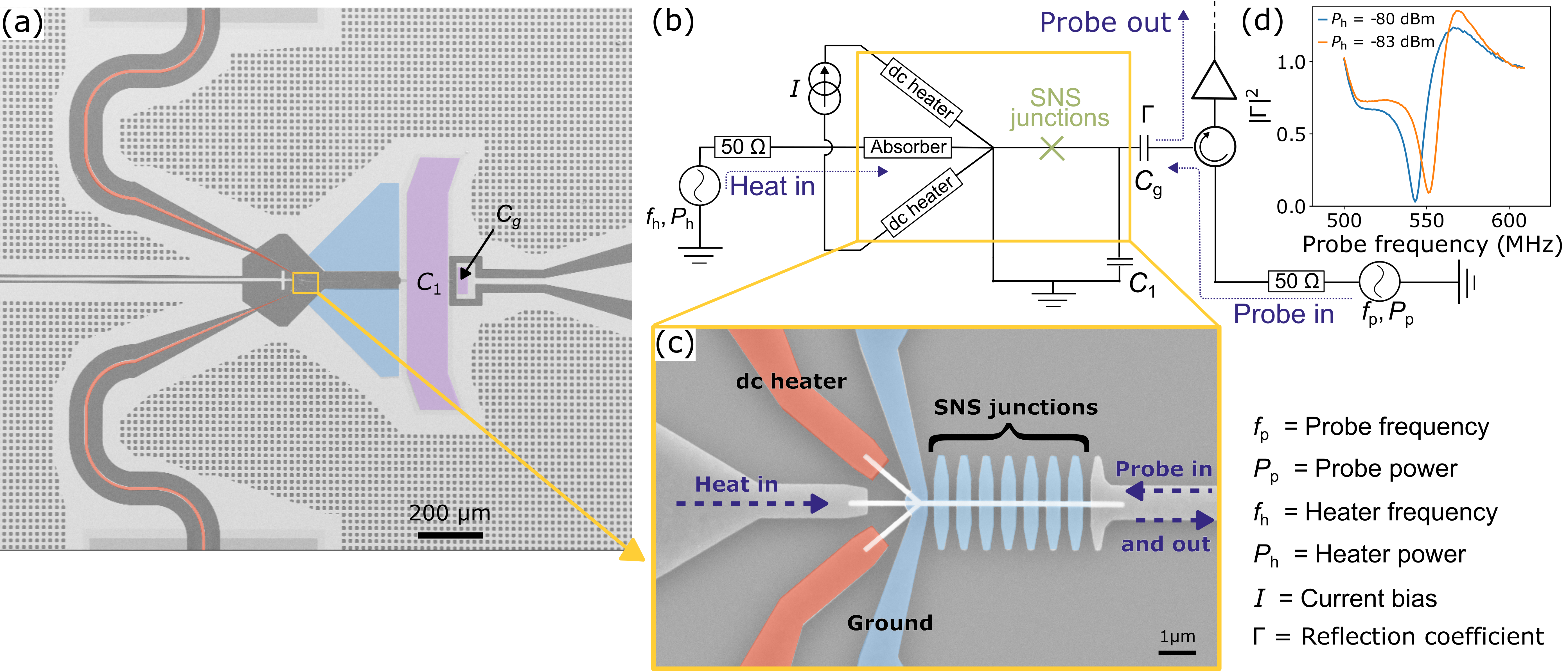}
    \caption{Description of the bolometer sample. (a) False-color optical image of the bolometer with a dc heater. The traceable electrical current is applied to the heater through a dc-heater line (red), and the sample is grounded through aluminum leads (blue). (b) Circuit diagram of the measurement setup indicating the shunt capacitor $C_1$ and gate capacitor $C_\textrm{g}$ also shown in (a) in purple. (c) False-color scanning-electron-microscope image of the detector. The long junctions operating as resistive power absorbers are 1-\textmu$\rm{m}$ long, and the short junctions operating as the thermometer are 300-nm long. The SNS junctions operating as the thermometer are formed by a AuPd nanowire (white) galvanically connected to superconducting Al islands and leads. The absorber is either heated by the microwave heater signal or by the dc heater. (d) Fraction of the probe power reflected at the gate capacitor as a function of probe frequency for heater powers of $-80\, \rm{dBm}$ (blue) and $-83\, \rm{dBm}$ (orange) at the absorber. An increase in the heater power induces a redshift of the resonance frequency.}
    \label{bolometer_scheme}
\end{figure*}
Historically, the radio frequency (rf) power has been defined by a substitution to electrical dc standards~\cite{1457679}, the Josephson voltage standard and the quantum Hall resistance standard, and therefore relies on calorimetric techniques. The International System of Units (SI) was redefined in 2018 through the fundamental physical constants, which has inspired traceable measurements of rf power using radiation pressure~\cite{Holloway_redefine_SI_power} and Rydberg atom sensors~\cite{Rydberg_atom}.
According to this redefinition, also the above-mentioned dc substitution provides a traceable path to the SI. It has been a common method for measuring microwave power in room temperature bolometers~\cite{HUBER20071327,Pradere_RoomT_bolometer}. Furthermore, the substitution has been utilized to calibrate low-temperature bolometers, for example, for X-ray detection~\cite{AHR1992387} and astronomical instruments~\cite{tucker1994cryogenic,10.1093/mnras/stt090}. In astronomy, detectors may be calibrated using atmospheric models and reference black-body calibration sources such as Mars~\cite{doi:10.1063/1.3637460}. In the optical regime where single-photon detection is relatively standard due to the high photon energy, the detectors can be calibrated using, for example, heralded photons~\cite{avella2011self}.
Bolometers used in particle detectors are typically calibrated using radioactive sources~\cite{arnaboldi2011characterization, cushman2017detector, abdelhameed2020new}. However, the traceable calibration technique we use in this manuscript is, in principle, applicable in all of the above cases.

Traceability to the SI at low power, typically below $-70$~dBm, may be challenging, as it falls below the operating power range of typical commercial power sensors~\cite{thesis_rf_sensors}.
Therefore, calibration of rf power in the cryostats widely used in quantum technology with heavily attenuated lines is inconvenient since the high attenuation calls for characterization in segments, yet the components are physically inaccessible when the cryostat is cold. 
In addition, if one has only a room temperature power meter, the signal has to be taken back from cryogenic temperatures to the room temperature. Since the lines are typically heavily attenuated, this requires amplification and calibration of the gain of the amplification chain, which adds uncertainties and noise to the measurement. For example for this reason, we recently introduced a method of calibrating the gain of cryogenic amplification chains~\cite{doi:10.1063/1.5096262}, but yet a cryogenic power sensor seems more appealing for measuring rf power at low temperatures.

Devices used in circuit quantum electrodynamics \cite{cQED_Blais,cQED_Wendin} (cQED) call for accurate characterization of the ambient radiation and their microwave properties at ultralow power levels. Experiments in cQED are based on superconducting circuits that use nonlinear elements, i.e., qubits, which operate mainly between 4 and 20 GHz \cite{qubit_review} and at a low signal level to prevent unwanted artifacts such as alternating-current (ac) Stark shifts~\cite{stark-effect_qubit}. 
In this context, there is a strong incentive to develop efficient and practical detectors operating in the microwave range.
Although some detectors for itinerant microwave photons recently managed to reach single-photon sensitivity with efficiencies up to~$96\%$ \cite{dassonneville2020numberresolved,Nakamura_mw_detector_2018,Wallraff_mw_detector_2018}, they rely on discrete qubit transitions or on cavity-confined photons to facilitate detection \cite{N_Roch_cavity,Wenner_cavity}. This limits signal amplitude calibration to a narrow relative bandwidth~\cite{photon_calib_Deppe} with the possibility of extending it to 1~GHz by observing a high-level ac Stark shift in a multilevel quantum system with a large frequency detuning \cite{stark_shift_calib, 2020_Schneider}, but this extension comes at the cost of reduced energy sensitivity. Another method~\cite{Calib_with_qubit} enables absolute calibration of power over a gigahertz-wide frequency range by measuring the spectra of scattered radiation from a two-level system in a transmission line. In addition, in situ characterization of qubit control lines in the megahertz regime has been implemented~\cite{in_situ_calib_line} using a transmon qubit coupled to a readout resonator. Analyzing the power dependence of the magnitude of the reflection coefficient at resonance of a tunable transmon-type superconducting circuit has also recently been employed as a radiation field thermometer~\cite{Scigliuzzo-PRX-2020}.  Overall, it is especially challenging to characterize ultralow microwave power at unknown frequencies.

Despite the great progress sped up by the race for a useful quantum computer, experiments in cQED usually lack tools to achieve reliable traceable power absorption measurements over a broad relative frequency band and over a range from fractions of photons to several photons in the frequency band of the signal. This tool gap hinders the accuracy and speed of key characterization measurements and hence introduces limitations for the operation characteristics of quantum-technological and other cryogenic devices.
Consequently, there is an urgent need for new tools to fill the gap in implementing the standard for ultralow rf power.
In this work, we introduce a traceable power absorption measurement based on dc substitution to enable high-precision measurements and furthermore provide a simple way to calibrate microwave power in a cryogenic environment.
For this purpose, we deem bolometers to be a promising candidate because of their naturally broadband low-temperature input, low operation power, small size, and simplicity. They can even achieve remarkable sensitivity, since recent studies suggest that ultralow-noise bolometers based on the Josephson effect may enable calorimetric measurement of itinerant microwave photons in the framework of cQED~\cite{Nature_roope,Efetov_nature_graphene}.

\section{Detection scheme}

\begin{figure}
    \includegraphics[width=8.6cm]{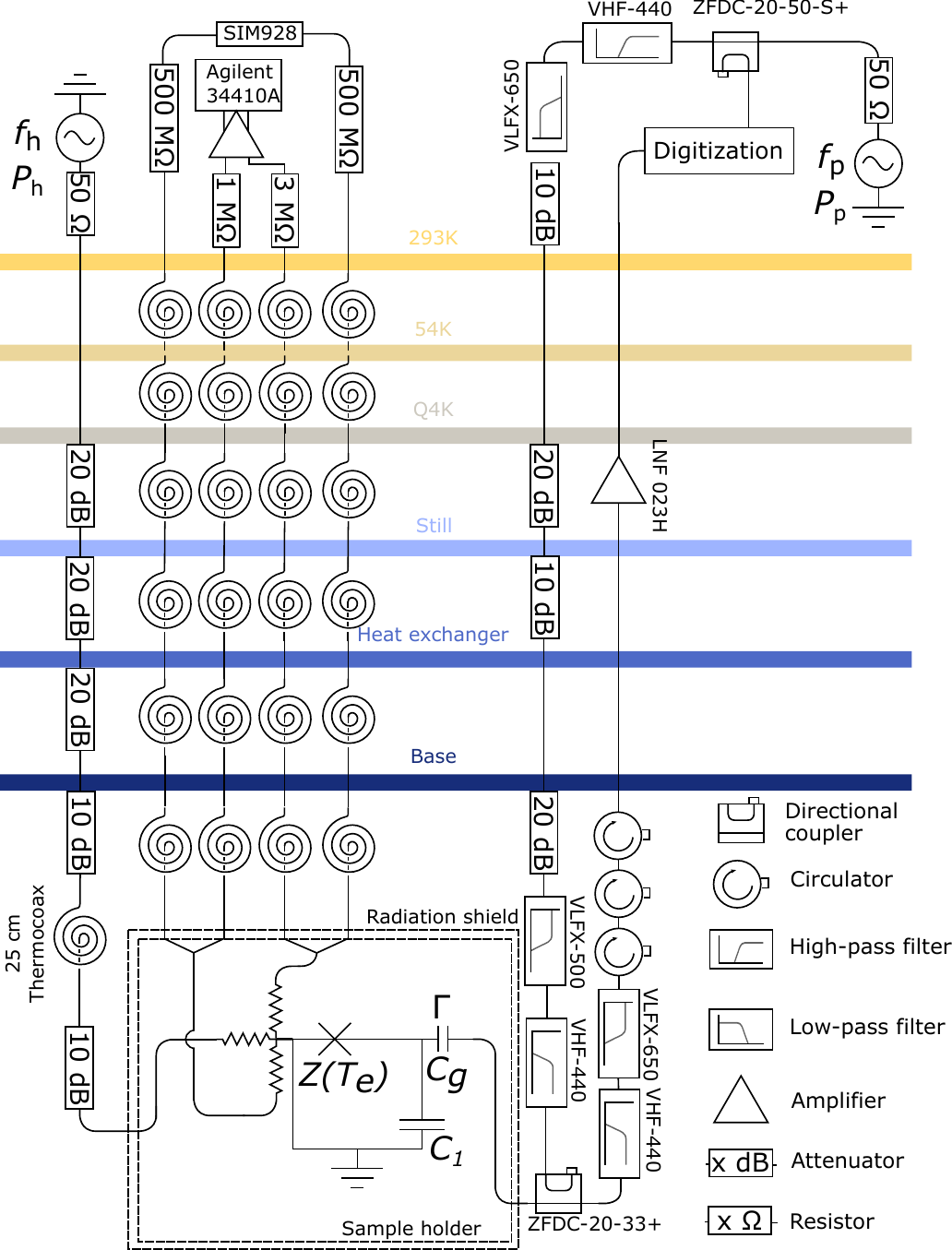}
    \caption{Main experimental setup for a traceable power absorption measurement with a bolometer equipped with a dc heater. The sample is placed inside a radiation shield to mitigate external noise. Thermalization of the lines going down to the base temperature of the cryostat is ensured by the presence of additional cable length (spiral shape) between each temperature stage.}
    \label{setup_scheme}
\end{figure}

The application of bolometers in the gigahertz regime~\cite{Pirro_review_bolometer} has required the development of nanoscale thermal absorber elements with reduced heat capacity achieved through optimization of both device geometry and materials~\cite{review-bolometer-Giazotto}. Therefore methods of electronic thermometry in mesoscopic devices are directly relevant for the thermal detection mechanisms~\cite{Luukanen_giazotto}. In particular, rf reflectometry of a circuit with a temperature-dependent impedance offers fast readout with low-back-action from the rf probe~\cite{Schmidt-APL-2003}.  

This article further describes a thermal detection approach that uses superconductor--normal-metal--superconductor (SNS) junctions as the temperature-dependent impedance 
\cite{Nature_roope,govenius2016detection,Joonas_bolometer,kokkoniemi2018nanobolometer}.  As the temperature of the normal-metal region of an SNS element increases, its Josephson inductance increases which is observed experimentally as the lowering of the resonance frequency of a tank inductance-capacitance (LC) resonator. In the device, electrical properties of the SNS junction are governed by the length and diffusion constant of the normal-metal weak link itself rather than the superconducting leads~\cite{dubos2001josephson, lake2017microwave, 2016_Ke_Graph_Josephson}. The minigap in the density of quasiparticle states is strongly temperature-dependent even for temperature excursions of a few millikelvin~\cite{Gueron_proximity_effect}. At frequencies corresponding to energies below the minigap, a low-power microwave signal can be used to probe the superconducting critical current in the SNS junctions. This quantity depends on the electron temperature. 

Other schemes have utilized measurements of the temperature-dependent switching current of a Josephson junction, to probe local heating of an SNS junction~\cite{Efetov_nature_graphene,Joonas_bolometer}. The method we describe in this article stems from an identical physical phenomenon even though we use an rf probe, i.e., the decrease in the critical current $I_\textrm{C}$ is equivalent to an increase in the Josephson inductance~\cite{tinkham} $L_\textrm{J}$, ideally as $I_\textrm{C} = (\Phi_0 / 2 \pi) / L_\textrm{J}$, where $\Phi_0$ is the magnetic-flux quantum. In addition, the real part of the junction admittance increases when temperature is increased but this effect is comparatively weak when the probe frequency and bath temperature remain below the minigap energy \cite{lake2017microwave,govenius2016detection,virtanen-prb-2011}. The devices are probed with an rf signal that pushes current through the junction at a level below its critical current, and hence the junction never switches to the normal-state during the measurement.

Figure~\ref{bolometer_scheme} shows our implementation of a bolometer for accurate traceable microwave power detection in the cryogenic environment. Additional details on sample fabrication are reported in Appendix \ref{app:sample_details}. The device has three distinct key components: (i) a thermal absorber, (ii) a readout circuit, and (iii) a dc heater, each of which we describe below. The central element of the bolometer is a continuous normal-metal--gold-palladium ($\mathrm{Au}_x \mathrm{Pd}_{1-x}$) nanowire that is shown in detail in Fig.~\ref{bolometer_scheme}(c). The nanowire is contacted by superconducting Al electrodes at various locations. The chemical composition of the normal-metal nanowire is reported in Appendix \ref{app:edx}.

In Fig.~\ref{bolometer_scheme}(c) the thermal absorber is comprised of the left section of nanowire that is longer than the coherence length of the superconducting aluminum. Therefore, the absorber is designed to be a matched resistor that can terminate a microwave transmission line with a characteristic impedance of $Z_{0} = 50\,\Omega$ to absorb rf power.  In the measurements that are described below, the microwave heater signal is provided by a room temperature signal generator and is delivered through  $50$-$\Omega$ coaxial transmission lines with thermalized filters and attenuators to the 10-mK base temperature stage of the cryostat. Once the signal reaches the sample, it is absorbed by the nanowire, increasing its temperature.  

In contrast to the absorber, the readout circuit of the bolometer [Fig.~\ref{bolometer_scheme}(a)--(c)] is made up of a series of eight shorter SNS junctions that assume reactive electrical behavior due to the presence of a minigap and finite Josephson inductance within each weak link.  In order to achieve readout with rf reflectometry, the series of eight SNS junctions is placed in parallel with an on-chip capacitance $C_{1}$ to ground, to realize an LC readout circuit resonating at $f_{r} \sim $~500~MHz. We probe the readout circuit using a low-power microwave probe so that the SNS junctions are well approximated as a linear inductor~\cite{govenius2016detection}. Since heating increases the inductance of the SNS junctions, the resonance frequency shifts down with increasing heating power as shown in Fig.~\ref{bolometer_scheme}(d). Importantly, the thermal absorber and the readout circuit are  electrically decoupled from one another by the central electrode that shunts the center of the nanowire to ground, i.e., the probe tone does not dissipate at the absorber and dc heater.  However, since both the probe and absorber sections are formed by a common nanowire, these sections are thermally connected and defined by a single temperature on the relevent timescales, but we note that temperature equilibration throughout the nanowire is not necessary when operating the dc substitution method described below. The correct operation requires only an identical response at the thermometer from the two different heating types. Based on an experimental study of metallic wires with similar dimensions and at similar temperatures~\cite{e-e_relaxation_time}, we expect the quasiparticles to quickly relax and reach thermal quasi-equilibrium at least within a nanosecond timescale. Additional discussion on the thermal conductance of the detector is reported in Appendix \ref{app:heat_losses}.




\begin{figure*}[t]
    \includegraphics[width=\textwidth]{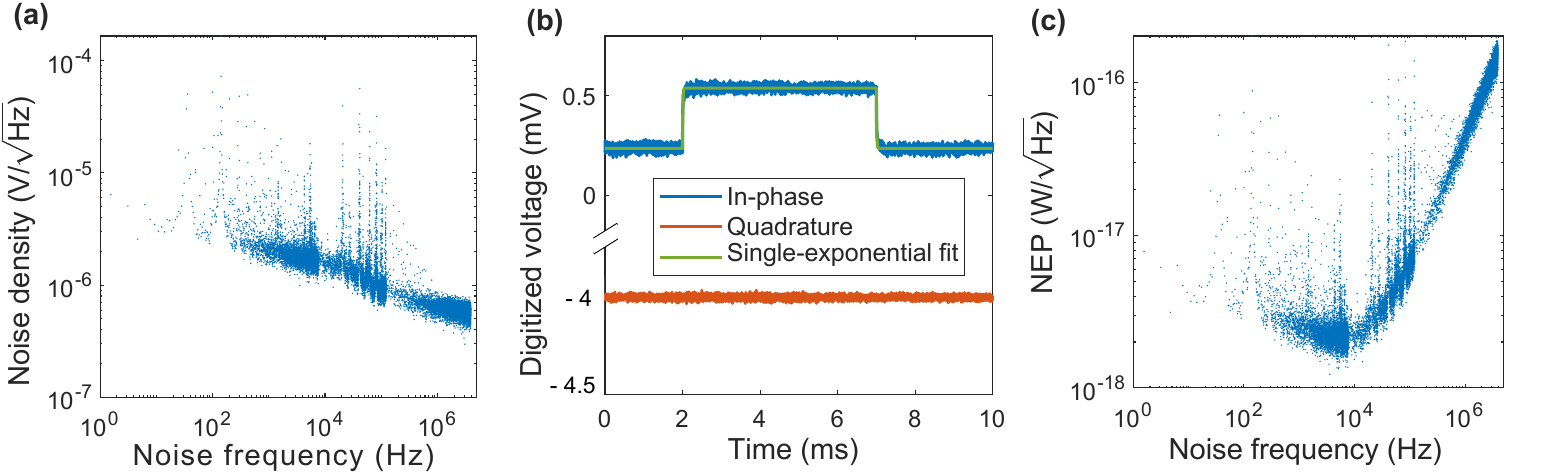}

    \caption{(a) Spectral density of voltage noise measured for the bolometer probe signal. (b) Time trace of the in-phase (blue) and quadrature (red) voltage of the measured probe signal for the heater power turned on at the time $t=2$ ms and turned off at $t=7$ ms. The green line denotes a single-exponential fit of the in-phase part. (c) Noise equivalent power (NEP) of the bolometer as a function of noise frequency. }
    \label{fig:NEP_timetrace}
\end{figure*}

Let us discuss the third component of the SNS bolometer which is introduced for the first time in this article, i.e., the dc heater that provides Joule heating locally to the bolometer for power calibration purposes. As shown in Fig.~\ref{bolometer_scheme}(c), we have redesigned the shape of the AuPd nanowire to split into thee long resistive SNS junctions, one operating as the usual rf absorber and two as the dc heater. The dc heater is galvanically connected to its superconducting leads enabling the heating of the nanowire with an accurately applied direct current.  

The calibration of the rf power absorbed at the device is described in detail in Secs.~\ref{sec_M1} and~\ref{sec_M2}. In a nutshell, we first apply a dc-heater current inducing heating on the order of a few femtowatts and observe the corresponding change in the transmission spectrum of the probe signal. Next, we sweep the rf heater power (and frequency) and record the response of the probe transmission.  With this data we can carry out a dc substitution method to calibrate the rf power delivered to the absorber by minimizing the residuals between the transmission spectrum under dc and rf heating. Importantly, we can establish a path to metrological traceability for the rf heating power even at extremely low levels because we can utilize the pre-calibration of the room temperature voltage source and of the resistors used to generate the dc-heater current.  

To measure microwave power at room temperature, a typical experiment using a low-temperature detector requires an accurate calibration of the attenuation of the rf line, through which the heater signal is applied. Figure~\ref{setup_scheme} shows a detailed diagram of our experimental setup. Multiple attenuators, amplifiers, filters, and directional couplers are placed between the sample and the instruments to enhance the signal-to-noise ratio. This setup enables us to accurately measure the power that is transmitted at frequency $f_{\textrm{h}}$ from room temperature through the cabling and into the broadband cryogenic power sensor at the base temperature of the dilution refrigerator. 
The specifications provided for the commercial devices in Fig.~\ref{setup_scheme} give a rough estimate of the line attenuation, but a more careful investigation is needed since many of the components are specified to work at or near room temperature. Attenuations can be measured with a vector network analyzer (VNA), but this can be difficult with heavily attenuated rf lines. Furthermore, VNA measurement necessarily includes attenuation, or gain, of the return path. This can be circumvented to some extent by measuring the gain of the return path with the so-called Y-factor technique.
Unfortunately, such a result is valid only for the specific setup corresponding to a typical uncertainty of 1~dB (26 \%) which occurs at a power level of $-130$~dBm (see supplemental material of Ref.~\citenum{joonas_zJ_biblio}). Furthermore, since thermal cycles may affect the components, such VNA calibration needs to be carried out at each cooldown, regardless of whether the setup was modified or not.

The presence of a dc heater at the bolometer enables us to bypass these limitations by carrying out detection in situ at the base temperature stage of the dilultion refrigerator. Instead of a full setup calibration, we only need to determine the response of the bolometer as a function of the dc heater power. Note that this method can be applied without changing the base temperature of the cryostat.
Therefore, our results reveal the amount of microwave power absorbed at the reference plane of a quantum device.

The $1\sigma$ uncertainty estimates of the different instrument calibrations and of the derived quantities are given in Table \ref{tab:errors}. We assume uncertainties to be independent, and hence we added them in squares for derived quantities.  Furthermore we transformed between the linear $(U_\textrm{lin})$ and logarithmic $(U_\textrm{log})$ uncertainties using the equation $U_\textrm{log} = 10\log_{10}(\textrm{e}) \times U_\textrm{lin}$~dB where $\textrm{e}$ is Euler's number.

\begin{table*}
\caption{\label{tab:errors} Uncertainty estimates ($1\sigma$) of the different instrument calibrations.}
\begin{ruledtabular}
\begin{tabular}{ ccc } 
 Instrument or derived quantity & Linear uncertainty (\%) & Logarithmic uncertainty (dB) \\ 
 \hline
 SIM 928 & 0.016 & 0.0007 \\ 
 Voltmeter & 0.003 & 0.0001 \\ 
 Amplifier gain @ 60 dB & 0.4 & 0.02 \\
 1-G$\Omega$ reference resistor & 0.04 & 0.0017\\
 1-G$\Omega$ resistor & 0.06 & 0.0026 \\
 Gigatronics 2550B @ 100~MHz& 9.6 & 0.42 \\
 dc heater power & 1.1 & 0.05
\end{tabular}
\end{ruledtabular}
\end{table*}


\begin{table*}[]
\caption{\label{tab:attenuation_err} Uncertainty estimates ($1\sigma$) for the bolometer readout, the power output of the microwave signal generator (Gigatronics 2550B), and the resulting measured attenuation for the black points in Fig.~\ref{fig:ssq}b.}
\begin{ruledtabular}
\begin{tabular}{rlll}
\multicolumn{1}{l}{}                        & \multicolumn{3}{c}{Heater frequency} \\
\multicolumn{1}{l}{}                        & 50 MHz     & 100 MHz     & 1 GHz     \\
 \hline
\multicolumn{1}{l}{Statistical uncertainty} &            &             &           \\
Linear (\%)                                 & 1.2        & 1.9         & 0.6       \\
Logarithmic (dB)                            & 0.05       & 0.08        & 0.03      \\
 \hline
\multicolumn{1}{l}{Gigatronics 2550B uncertainty} &            &             &           \\
Linear (\%)                                 & 1.6        & 9.6         & 8.4       \\
Logarithmic (dB)                            & 0.07       & 0.42        & 0.36      \\
 \hline
\multicolumn{1}{l}{Attenuation uncertainty}       &            &             &           \\
Linear (\%)                                 & 2.29       & 9.88        & 8.31      \\
Logarithmic (dB)                            & 0.1        & 0.43        & 0.36     
\end{tabular}
\end{ruledtabular}
\end{table*}

%

\begin{figure*}
    \includegraphics[width=13cm]{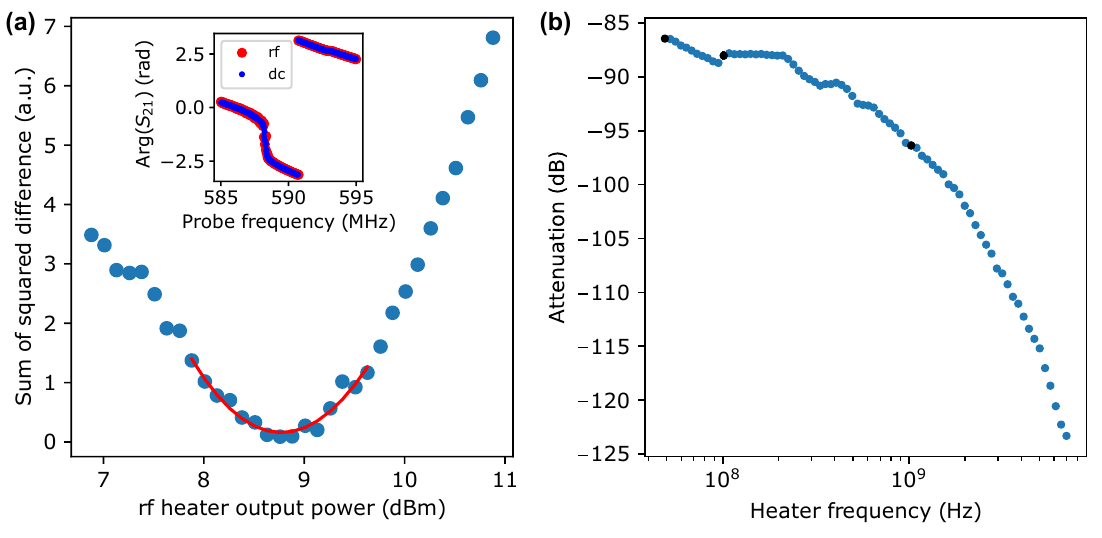}
    \caption{Comparison of transmission curves used to deduce the equivalence between dc and absorbed rf heating powers. (a) Example of the sum of squared differences of the complex-valued transmission coefficients $S_{21}$  as a function of the applied rf heater power generated by a signal generator at a 7-GHz frequency for an applied dc heater power of $-114.03$~dBm. The minimum of the fitted third-order polynomial function is achieved at 8.77~dBm. The attenuation at 7~GHz is $-122.80$~dB, of which 0.55~dB arises from the imperfect matching of the rf absorber impedance to the 50-$\Omega$ transmission line. Inset: Phase of $S_{21}$  as a function of probe frequency for dc and rf heating and 8.75~dBm of applied rf heating. (b) Attenuation of the rf-heater line as a function of the heater signal frequency. The strong frequency dependence is due to a Thermocoax cable used in the setup. The black markers indicate discrete frequencies where output power of the microwave source was calibrated using a commercial power meter. The uncertainty of the attanuation at the black markers is provided in Table.~\ref{tab:attenuation_err}. The data corresponding to the blue markers lacks the calibration of the source and has hence a greatly elevated uncertainty.}
    \label{fig:ssq}
\end{figure*}

\section{Measurement}

%


To characterize our cryogenic power sensor in more detail, we define and measure the noise equivalent power (NEP) as in our previous work \cite{Roope_JPA_biblio} by first measuring the noise spectrum of the output and then dividing it by the quasistatic responsivity of the detector. We further take into account the frequency dependence of the responsivity by dividing the quasistatic responsivity by a factor $\sqrt{1+(2\pi f_\textrm{n}\tau)^2}$, where $\tau$ is the measured time constant of the detector, and $f_\textrm{n}$ is the noise frequency. This procedure yields the noise of the bolometer output in units of the input power, which we take as the definition of NEP. The experimental data shown in Figure \ref{fig:NEP_timetrace} indicate that the lowest measured NEP is roughly $2\, \textrm{aW}/\sqrt{\textrm{Hz}}$, together with a 20~{\textmu}s thermal time constant. For power sensing, the thermal time constant suggests a maximum repetition rate of the order of ten kilohertz. To measure, for example, an incoming power of 10~fW at a relative uncertainty of 1\%, one needs an integration time of 0.2~ms according to the measured NEP, which is not limited by the thermal time constant. 
The measured NEP is roughly two orders of magnitude greater than that previously reported \cite{Roope_JPA_biblio} on similar sample without a dc heater. This difference can be explained by the larger volume of the metal in the dc bolometer and, perhaps more importantly, by the presence of two dc lines inducing additional noise channel for the sample.


Using Method I or II described below, one needs to carry out several measurements of the reflection coefficient to measure the attenuation of the cabling. In our current experiment we sweep the rf heater power and compare it to a single dc heater power.
Alternatively, to determine the power of an unknown signal, one can use calibration data for the reflection coefficient at different dc heater powers. Note that if one has obtained beforehand such calibration data, then the total measurement time is limited to the acquisition of a single trace of the transmission coefficient.

\section{Results: Method I}
\label{sec_M1}

In the following, we present in detail our calibration method.
First, we measure over a range of probe frequencies $f_{\mathrm{p}}$ the transmission coefficient $S_{21}$ of the probe signal, from the source to the digitization stage shown in Fig.~\ref{setup_scheme}, with finite dc heating such that the resonance shifts by approximately 1~MHz with respect to the zero-bias case. Next, we repeat the measurement at zero dc-heater power but at a considerable output power of the rf heater signal generator and calculate the sum of the squared differences between the rf-heating trace $S_{21}^{\mathrm{rf}}(f_{\mathrm{p}})$ and the dc-heating $S_{21}^{\mathrm{cd}}(f_{\mathrm{p}})$, i.e., we define the residual as
\begin{widetext}
\begin{equation}
R_0 = \sum_{f_{\mathrm{p}}}^{} \big( \{\mathrm{Re}[S_{21}^{\mathrm{dc}}(f_{\mathrm{p}})]-\textrm{Re}[S_{21}^{\mathrm{rf}}(f_{\mathrm{p}})]\}^2 + \{\mathrm{Im}[S_{21}^{\mathrm{dc}}(f_{\mathrm{p}})]-\mathrm{Im}[S_{21}^{\mathrm{rf}}(f_{\mathrm{p}})]\}^2 \big).
\label{eq_ssq}
\end{equation}
\end{widetext}
A small residual indicates nearly equal absorbed power induced by the rf and dc heating sources.
In Fig.~\ref{fig:ssq}(a), we show the residual as a function of the applied rf power to the heater line.
We find the rf power corresponding to the dc-heater power by fitting a third-order polynomial to the data near the minimum. We repeat such measurements many times to estimate the statistical uncertainty of the measurement. 

\begin{figure*}[t]
   \includegraphics[width=\textwidth]{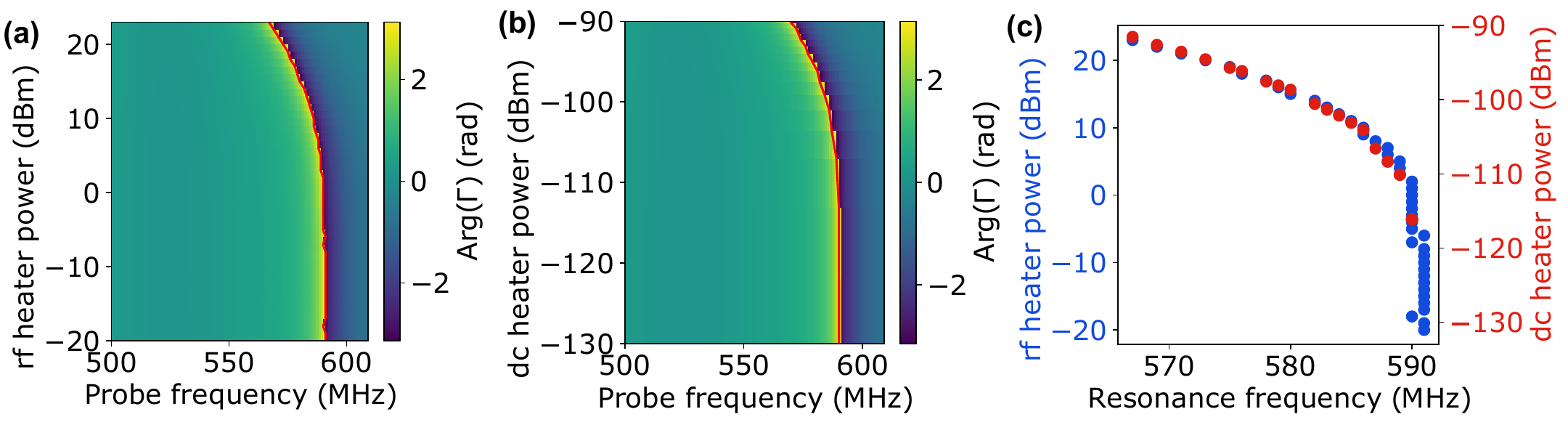}
    \caption{Deduction of the attenuation of the heater line with Method II. Phase of the reflection coefficient $\Gamma$
    as a function of the probe frequency and the (a) rf and (b) dc-heater power. The resonance frequency is depicted by the red line. (c) Comparison of the resonance frequencies obtained by the dc and rf heating at $f_{\mathrm{h}} = 5\, \mathrm{GHz}$. In this example, the best match between the two curves is obtained when considering $-114.3$~dB attenuation along the rf-heater line.}
    \label{fig:res_freq}
\end{figure*}

As expected, we observe in Fig.~\ref{fig:ssq}(b) a strong frequency dependency in the attenuation of the heater line due to the presence of the 25-cm Thermocoax cable\cite{doi:10.1063/1.1145385}.
In addition, we have frequency-dependent attenuation arising from the other coaxial cables in the heater line illustrated in Fig.~\ref{setup_scheme}. However, at low frequencies where the coaxial conductor losses are small, we find that the attenuation approaches 80~dB, which corresponds to the cumulative nominal value of the added attenuation present in the setup. The 1$\sigma$ confidence interval of the type A uncertainty of the measured attenuation is below 0.1~dB for approximately 10~min of measurement time. Figure~\ref{fig:ssq}(b) indicates the discrete frequencies (50~MHz, 100~MHZ, and 1~GHz) where the output power of the Gigatronics room temperature microwave generator was calibrated using a commercial room-temperature microwave power meter (see Appendix~\ref{app:cal} and Table~\ref{tab:attenuation_err}). Below, we estimate the uncertainty of the measured attenuation at these frequencies. Outside these frequencies, we do not give a precise estimation of the uncertainty, which is likely of the order of a few decibels.

In addition to the noise in the measurement, we have systematic error arising from the calibration of our instruments. The current bias is provided by a Stanford Research Systems SIM~928 voltage source with nominal $0.5$-G$\Omega$ resistors in both dc lines. The total measured resistance is $0.9955\, \textrm{G}\Omega \, \pm \, 0.6 \, \textrm{M}\Omega$. The SIM voltage source has an uncertainty of 160 ppm. The dc lines and the on-chip dc heater resistors have resistances well below 1 k$\Omega$. The effect of the resistors on the bias current can be neglected without loss of accuracy. 

We measured the resistance of the dc-heater resistors in a four-probe configuration using an Agilent digital multimeter model 34410A as a voltmeter. The voltage was amplified with a Femtoamp DLPVA-100 voltage amplifier, and the nominal 60-dB gain was measured to be $60.02\, \textrm{dB} \pm 0.0013 \, \rm{dB}$. 
The resistance of the heater $R_{\rm{heater}}$ determined from the slope of the measured current-voltage curve yields $(48.8 \pm 0.5) \Omega$. 
The dc power reaching the bolometer has an uncertainty of approximately 0.05~dB, which is limited by noise and the uncertainty of the dc-heater resistance.

Consequently, we observe that at 100~MHz, the uncertainty of the measured attenuation in Fig.~\ref{fig:ssq}(b) is clearly dominated by that of the room temperature microwave source, 0.33~dB. At 50~Mhz and 1~GhHz, the uncertainty of the attenuation is greatly improved owing to only 0.01-dB contribution from the microwave source, and hence it is dominated by the 0.1~dB statictical uncertainty of the measurement.


We observe several unexpected features in Fig.~\ref{fig:ssq}(b) such as plateaus with kinks and a sharp drop of the attenuation just before 100~MHz. We attribute these to the room temperature microwave source which is not accurately calibrated at these frequencies. The drop in the attenuation is likely reminiscent of a drop of roughly 2~dB in the output power of the microwave source in the vicinity of 100~MHz. 
In the future, the accuracy of the attenuation measurement in the whole frequency range can be conveniently improved by a fully calibrated microwave source. However, we note that the source uncertainty only affects the attenuation measurement, not the measurement of the absorbed rf power at cryogenic temperatures. Further calibration details are provided in Appendix~\ref{app:cal}.

\section{Results: Method II}
\label{sec_M2}
Alternatively, we can calibrate the rf heater power by measuring the resonance frequency at multiple dc-heater currents and output powers of the rf-heater signal generator as shown in Fig.~\ref{fig:res_freq}. Finding the offset between the dc and rf heater powers that minimizes the difference between the resonance frequencies measured with the two heating methods yields the attenuation of the heater line. 
A similar method can be applied to quantities other than the resonance frequency. For example, we can measure the transmission coefficient at a given frequency as a function of dc and rf heater powers. In this case, the attenuation of the rf line is the offset power that minimizes the difference between the two transmission coefficients. However, we deem these alternative methods to be less robust since they do not utilize the full information on the resonance. Nevertheless, such a method could be greatly faster than the one we use above, and remains an appealing line of future research. \\

From Fig.~\ref{fig:res_freq}(c) we can also obtain an indication of the dynamic range of our power sensing technique using Method II. For the configuration given by the wiring of Fig.~\ref{setup_scheme}, the dynamic range exceeds 20~dB.  The measurement of the dynamic range is limited from above by the maximum output power of the room temperature Gigatronics source owing to the high total line attenuation. Otherwise, the trend of decreasing resonance frequency of the readout with increasing heating power could be extended further by increasing the microwave power delivered to the bolometer. Therefore, Fig.~\ref{fig:res_freq}(c) highlights a potential advantage of thermal detection with respect to qubit-based photon detection schemes discussed in Sec.~\ref{sec:intro}, i.e., the bolometer has the ability to resolve both single-photon power levels and higher input powers.

\section{Conclusions}
In summary, we developed a sensitive bolometer device for traceable microwave power absorption measurements relying on dc substitution. The device operates at low temperatures, exhibits a broad input bandwidth and is suitable for characterization of devices operating in the framework of cQED.
As an illustration of the utility of the introduced microwave sensor, we demonstrated the calibration of a heavily attenuated rf line including several microwave components at low temperatures. We achieved this by comparing the response of the bolometer to a heater signal applied through the rf line and to heating applied through a dc line. From this comparison, we accurately determined the frequency-dependent attenuation of the rf line with an uncertainty of 0.33~dB for an absorbed power of $-114$~dBm. We note that this method potentially leads to a substantial time reduction in setup characterization compared with the so-called Y-factor method that requires one to change the temperature of a resistor and consequently to wait for potentially slow saturation of the thermal relaxation.
This work aims to facilitate the implementation of bolometers in experiments on cryogenic electronics as traceable power sensors and to enable highly accurate power measurements.
From the noise performance of the device, we observed that the sample would benefit from improved shielding from dc line noise, and in future experiments, we plan to use such a bolometer for microwave signal calibration in qubit experiments.


\section*{Acknowledgments}
We acknowledge the provision of facilities and technical support by Aalto University at OtaNano – Micronova Nanofabrication Center and LTL infrastructure which is part of the European Microkelvin Platform (EMP, No. 824109 EU Horizon 2020).
We have received funding from the European Research Council under Consolidator Grant No. 681311 (QUESS) and under Advanced Grant Nos. 670743 (QuDeT) and 101053801 (ConceptQ), European Commission through H2020 program projects QMiCS (grant agreement 820505, Quantum Flagship), the Quantum Flagship funding by the European Commission through project No. 101113946 (OpenSuperQPlus100), the Academy of Finland through its Centers of Excellence Program (project Nos. 312300, 312059, and 312295) and grants (Nos. 336810, 314447, 314448, 314449, 305237, 316551, 308161, 335460, and 314302), the Finnish Cultural Foundation, the Vilho, Yrjö and Kalle Väisälä Foundation of the Finnish Academy of Science and Letters, the Jane and Aatos Erkko Foundation, and
the Technology Industries of Finland Centennial Foundation. 

\section*{AUTHOR DECLARATIONS}
\subsection*{Conflict of Interest}
M.M. declares that he is a Co-Founder and Chief Scientist of the Quantum-Computer company IQM. R.E.L.. R.K., and M.M. declare that they are inventors in a related family of patents titled \emph{Cryogenic microwave analyzer} (patent application number FI20176051A and herein) owned by IQM Finland Oy. The other authors have no conflicts to disclose.

\section*{DATA AVAILABILITY}
The data and analysis scripts are openly available on Zenodo at \url{https://doi.org/10.5281/zenodo.7568109}.

\appendix
\section{Sample details}
\label{app:sample_details}
\subsection*{Fabrication}
We begin with a high-resistivity ($\rho$  > 10 k$\Omega$ cm) 4-inch silicon wafer (100) covered by a 300-nm thermal oxide $\mathrm{SiO_{x}}$ and sputter on it 200 nm of pure Nb. We define the waveguide in a Karl Suss MA-6 mask aligner using AZ5214E photoresist in positive mode with hard contact. After development, the sample is etched with Plasmalab 80Plus Oxford Instruments reactive ion etching (RIE) system. The plasma works with a gas flow of $\mathrm{SF_{6}}$/$\mathrm{O_{2}}$ at 40 sccm/20 sccm with a rf-field power of 100 W.
After etching, the resist residuals are cleaned in an ultrasonic machine with acetone and IPA and dried with a nitrogen gun. Next, a thin film dielectric layer  $\mathrm{Al_{2}O_{3}}$ (45 nm) is grown by atomic-layer deposition (ALD) in a Beneq TFS-500 system. The dc-heater dielectric layer is protected using AZ5214E resist and the rest of the ALD oxide is wet-etched with an ammonium fluoride–hydrofluoric acid mixture. 
Then, the 4-inch wafer is cleaved into a 2 $\times$ 2 cm$^{2}$ chip by Disco DADdy. Subsequently, the nanowire is patterned by EPBG5000pES electron beam lithography (EBL) with a bilayer of MMA/PMMA resist on a single chip. The 30-nm-thick AuPd layer is deposited in e-beam evaporator at a rate of 0.5 Å/s. After liftoff in acetone overnight, the superconducting leads gavanically connected to the nanowire are patterned by EBL and deposited with 100 nm Al at a rate of 5 Å/s. Finally, each pixel (5 $\times$ 5 mm$^{2}$) is cleaved by a laser micromachining system and packaged with Al bonding wires.\\

\subsection*{Nanowire resistance}
\begin{figure}[b]
    \includegraphics[width=\columnwidth]{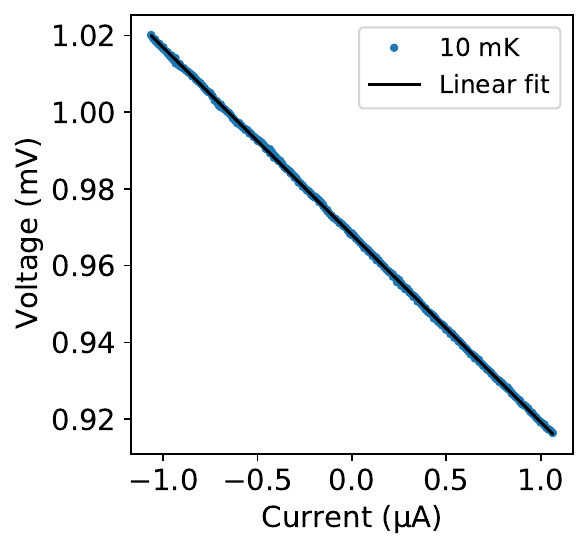}
    \caption{Voltage drop across the dc heater consisting of two AuPd resistors as a function of the applied current at 10 mK.}
    \label{fig:IV}
\end{figure}

Figure~\ref{fig:IV} shows the result of a four-probe measurement for the series resistance of the two dc-heater resistors at 10 mK with a two-second integration time and five averages. From a linear fit to the data, we obtain a resistance of 48.8 $\Omega$.

\section{Nanowire chemical composition}
\label{app:edx}
\begin{figure}[h]
\includegraphics[width=\columnwidth]{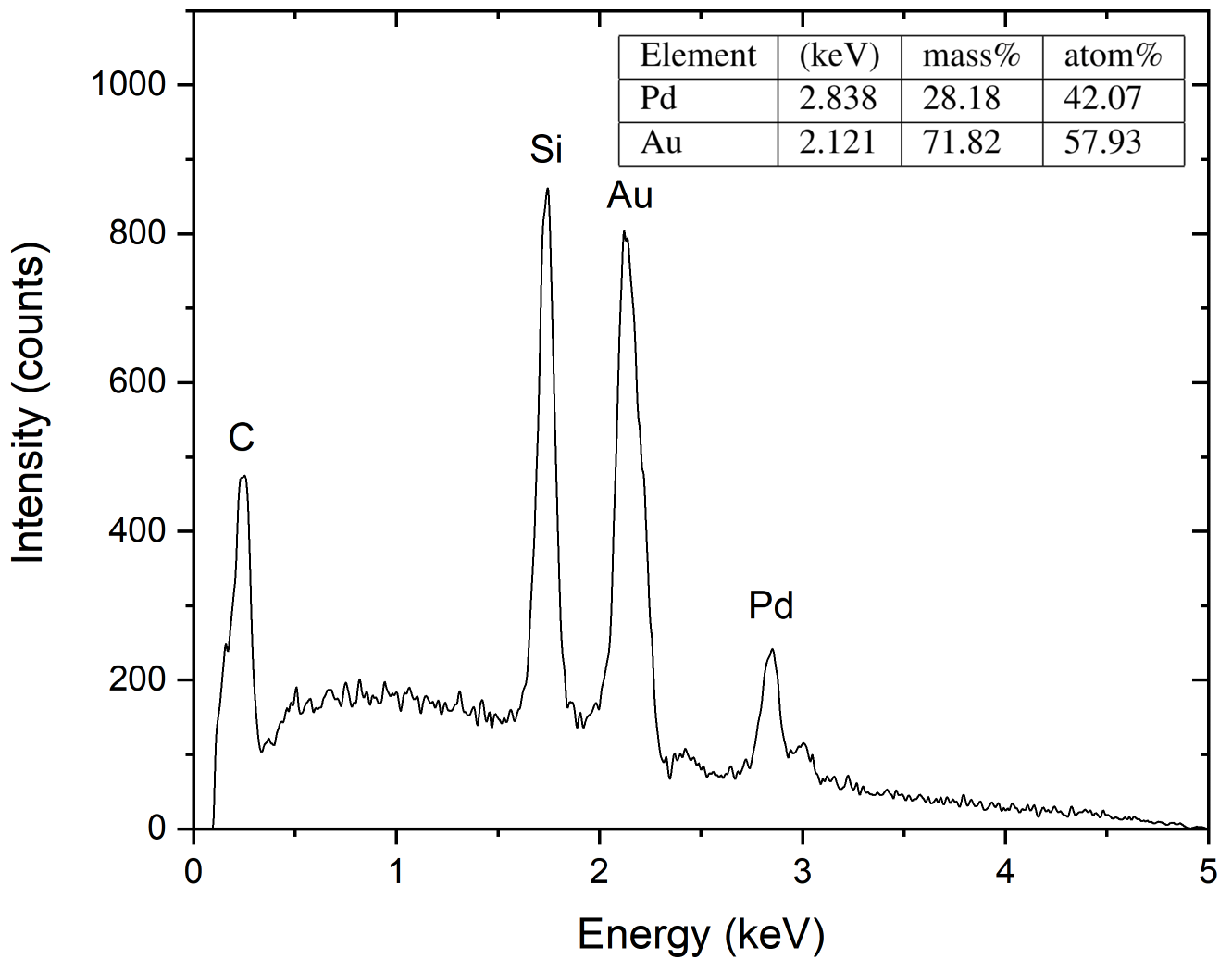}
\caption{Energy-dispersive X-ray spectroscopy of the AuPd nanowire of a nominally identical sample to that cryogenically measured in this work.}
\label{EDX}
\end{figure}

We evaluate the chemical composition of the AuPd nanowire by energy dispersive X-ray (EDX) analysis. We find the nanowire composition to be $\mathrm{Au}_{x}\mathrm{Pd}_{1-x}$ with $x\approx 0.58$.
To prevent damaging the measured sample this study is carried out on different but nominally identical samples fabricated in a different batch.
Figure \ref{EDX} shows the X-ray spectrum. The peaks at 2.121~keV and 2.838~keV correspond to Au and Pd, respectively. The electron beam energy is set to 5~keV during analysis.

\section{Heat losses and thermal conductance}
\label{app:heat_losses}
Similar to Ref.~\citenum{thermal_conductance_SNS}, one can estimate the thermal conductance in the SNS junctions~\cite{proximity_thermal_conducance} $G_{\mathrm{SNS}}$ and compare it to the thermal conductance to the cryostat phonon bath $G_{\mathrm{b}}$. To quantify this heat transfer, we consider the ratio $l_{\mathrm{s}}=L_{\mathrm{s}}/\xi_0$, where $L_{\mathrm{s}}$ is the length of the superconducting aluminum between the absorber resistor and the thermometer part of the bolometer and  $\xi_0$ is the the superconductor coherence length. We obtain a ratio of $l_{\mathrm{s}}\approx 1$ by considering $\xi_0 = \sqrt{\hbar D_{\mathrm{s}}/\Delta_0}$, where the Al bulk energy gap at zero temperature $\Delta_0 = 200\, \mathrm{\upmu eV}$, the diffusion constant of the superconductor $D_{\mathrm{s}} = 50\, \mathrm{cm^2/s}$, $L_{\mathrm{s}} = 300\, \mathrm{nm}$, and $\hbar$ is the reduced Planck constant.
According to Ref.~\citenum{thermal_conductance_SNS}, under these conditions $G_{\mathrm{SNS}}$ can be approximated by the Wiedemann–Franz value $\mathcal{L}_0 G_{\mathrm{N}} T$, with $\mathcal{L}_0$ the Lorentz number and $G_{\mathrm{N}}$ the normal state electrical conductance. This yields a thermal conductance of $G_{\mathrm{SNS}} = 0.4$~$\mathrm{nW/K}$, which is more than five orders of magnitude larger than the thermal conductance from the nanowire to the cryostat phonon bath that we previously reported on an otherwise similar device but without the dc heater~\cite{govenius2016detection}. Therefore, we expect the chain of SNS junctions to be strongly thermally coupled to the rf absorber and dc heater. 

\section{Calibration of instruments}
\label{app:cal}
The voltmeter and the amplifier were calibrated against a Fluke 5440B voltage calibrator, which in turn was shortly before the experiments calibrated by VTT MIKES in a way traceable to the Josephson voltage standard. Next, a calibrated voltmeter was used to verify the output voltage of the SIM 928 voltage source. The resistances of the bias resistors were measured with an Agilent 34410A multimeter using a four-probe configuration. The multimeter resistance reading was calibrated against the Measurements International model 4310H resistance standard. An Agilent N1913A power meter was used to verify the output power of the Gigatronics 2550B signal generator. The power meter was calibrated at VTT MIKES within a few weeks prior to the measurements reported in this article. \\


\section*{References}
\bibliography{main}

\end{document}